\documentclass[preprint]{aastex}
\usepackage{natbib,psfig}

\newcommand{\oviii}{O~{\small VIII}\,}
\newcommand{\ovii}{O~{\small VII}\,}
\newcommand{\nex}{Ne~{\small X}\,}
\newcommand{\neix}{Ne~{\small IX}\,}
\newcommand{\mgxii}{Mg~{\small XII}\,}
\newcommand{\mgxi}{Mg~{\small XI}\,}
\newcommand{\sixiv}{Si~{\small XIV}\,}
\newcommand{\fexxvi}{Fe~{\small XXVI}\,}
\newcommand{\fekalpha}{Fe~{\small K$\alpha$}\,}

\begin{document}


\title{High-Resolution X-Ray Spectroscopy of the Accretion Disk Corona Source 4U~1822$-$37} 
\author{Jean Cottam, Masao Sako,
        Steven M. Kahn, and Frits Paerels} 
\affil{Columbia Astrophysics Laboratory and Department of Physics,
       Columbia University, 550 West 120th Street, New York, NY, 10027} 

\and
\author{Duane A. Liedahl } 
\affil{Department of Physics and Advanced Technologies, Lawrence Livermore National Laboratory,
       P.O. Box 808, L-41, Livermore,  CA  94550} 

\received{}
\revised{}
\accepted{}

\shorttitle{High-Resolution X-Ray Spectroscopy of 4U~1822$-$37}
\shortauthors{Cottam et al.}

\begin{abstract}
We present a preliminary analysis of the X-ray spectrum of the accretion disk corona source, 4U~1822$-$37, obtained with the High Energy Transmission Grating Spectrometer onboard the {\it Chandra} X-ray Observatory.  We detect discrete emission lines from photoionized iron, silicon, magnesium, neon, and oxygen, as well as a bright iron fluorescence line.  Phase-resolved spectroscopy suggests that the recombination emission comes from an X-ray illuminated bulge located at the predicted point of impact between the disk and the accretion stream.  The fluorescence emission originates in an extended region on the disk that is illuminated by light scattered from the corona.  

\end{abstract}

\vspace{0.1cm}


\section{Introduction}
\label{sec:intro}

X-ray spectroscopy of low-mass X-ray binaries (LMXBs) provides quantitative constraints on the complex and poorly understood process of mass transfer through accretion disks.  The discrete spectral features that are produced as the central continuum emission propagates through the circumsource material are sensitive to the physical conditions and geometry of the accreting plasma.  High-resolution spectroscopy has been shown to be a powerful tool in empirically constraining the accretion process in LMXBs;  analysis of the spectra obtained with the {\it Chandra} and {\it XMM-Newton} observatories have provided quantitiative measurements of the plasma parameters and accretion geometries in Cir X-1 \citep{BrandtSchulz} and EXO~0748$-$67 \citep{Cottam}.  

In this letter we present a preliminary analysis of the 4U~1822$-$37 spectra obtained with the {\it Chandra} High Energy Transmission Grating Spectrometer (HETGS).  The smoothly-modulated, nearly color-independent partial eclipse ($P=5.57\, {\rm hr}$) and the unusually hard X-ray spectrum of this source led \citet{White81} and \citet{WhiteHolt} to conclude that the central source is surrounded and partially obscurred by an accretion disk corona, formed from material evaporated off of an X-ray illuminated disk.  Extensive modeling of the lightcurves by \citet{WhiteHolt}, \citet{MasonCordovaB}, and \citet{HellierMason} constrained the geometric parameters of the system ($i=75-85^\circ$, $R_{\rm disk}=5-7.3 \times 10^{10}\, {\rm cm}$, and $R_{\rm corona}\sim 2.5-4 \times 10^{10}\, {\rm cm}$), and suggested the presence of a phase-dependent vertical structure along the edge of the disk.  The X-ray spectra obtained with previous low-resolution instruments have been extremely difficult to model; the data require a complex continuum and a broad Fe~{\small K} line, however broad residuals at soft X-ray energies suggest the presence of additional unresolved emission or absorption structure \citep{White81,WKA,Parmar}.  

With the high spectral resolution of the HETGS we are able to resolve discrete, narrow line emission from a range of abundant ions.  As we describe below, detailed analysis of these features and their orbital phase variations suggest the presence of photoionized material on the inside edge of a bulge located at the point of impact with the accretion stream, and fluorescing material in the inner disk illuminated by scattered emission from the extended central corona. 
\\

\section{Data Reduction}
\label{sec:reduction}

4U~1822$-$37 was observed on 2000 August 23 with the {\it Chandra} HETGS \citep{Canizares} in the standard ACIS-S-HETG configuration.  The observation began at 16:20:37~UT and lasted for approximately $40\, {\rm ks}$ covering two full binary periods.  The source is bright with a phase-averaged flux of $4.8 \times 10^{-10}\, {\rm ergs/s/cm^{2}}$ in the $0.7-10\, {\rm keV}$ band.  The zero order image is therefore severely piled-up, making it unusable for spectroscopy.  However, in the combined positive and negative first order MEG and HEG spectra the phase-averaged count rates are only $2.8\, {\rm ct/s}$ and $2.0\, {\rm ct/s}$ respectively, and the spectra show no evidence of pile-up problems.  

We used the Chandra Interactive Analysis of Observations software (CIAO) to extract the spectra and generate the aspect-corrected effective area curves.  However, due to limitations with the CIAO 1.1 version that was available the time, we were unable to generate the necessary files for time intervals within the observation.  Instead we extracted the spectra for specific phases using custom-built software that was calibrated with the publicly available Capella data \citep[see][]{Behar}.  In order to utilize the latest effective area models, all global fitting was performed using the CIAO products within XSPEC \citep[version 11.0:][]{Arnaud}.  The phase-dependent analyses were performed using our own software.  For both procedures the positive and negative first order spectra were combined to maximize the statistical quality of the data.     
\\

\section{Observational Features}
\label{sec:features}

\subsection{Discrete Emission}
\label{subsec:discrete}

\begin{figure} 
      \centerline{\psfig{figure=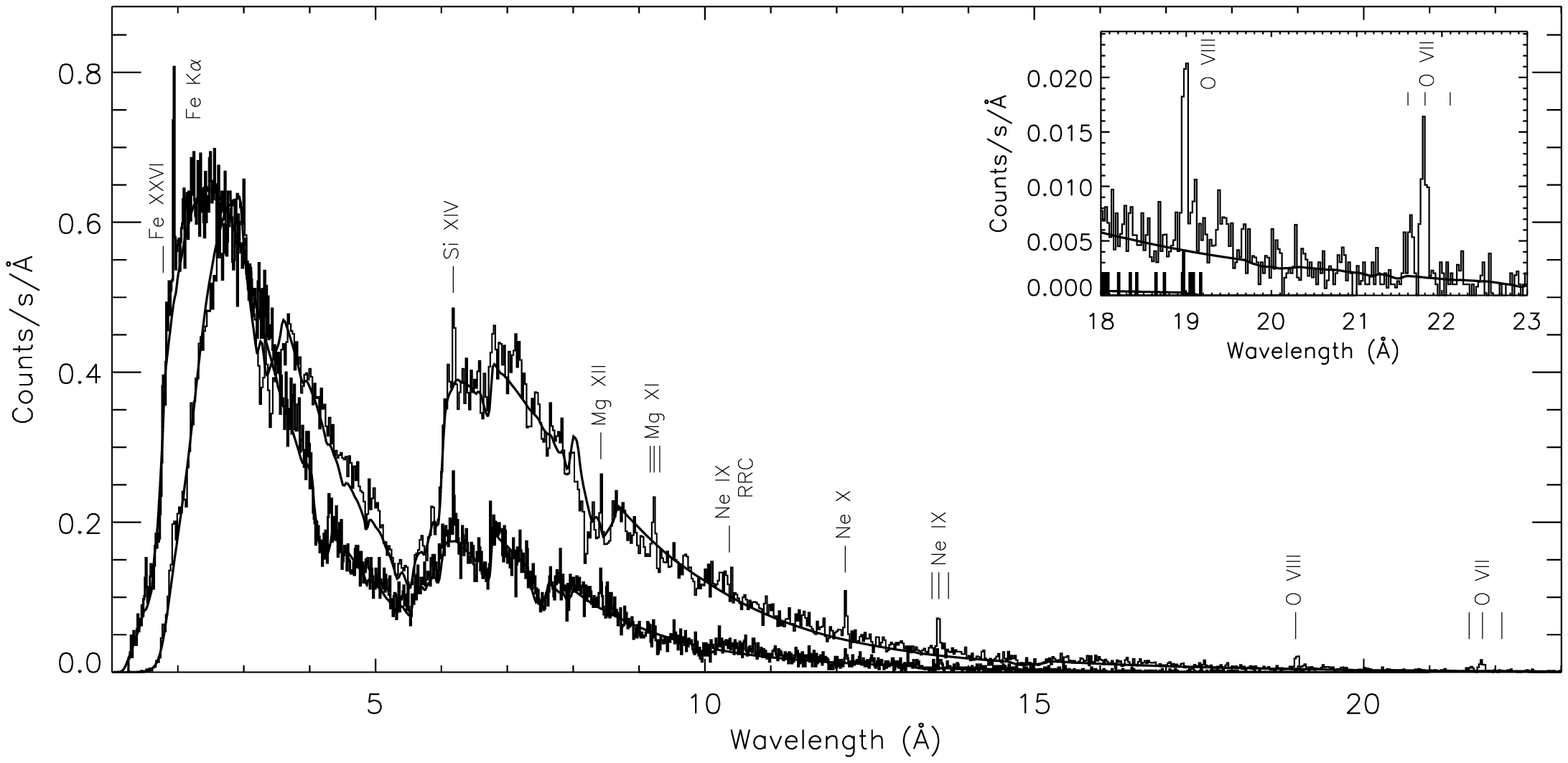,angle=90,height=5.5in}}
      \vspace{-4.5cm}
      \caption{HETGS spectra of 4U~1822$-$37.  The thin and thick histograms correspond to the MEG ($m=\pm 1$) and HEG ($m=\pm 1$) spectra, respectively.  The solid lines represent a simple fit to the continuum emission and are intended to indicate the instrumental features in the effective area.  } 
      \label{FigSpect}
   \end{figure}

In Figure ~\ref{FigSpect} we show the first order MEG and HEG spectra for 4U~1822$-$37 integrated over the total observation.  Discrete emission lines are clearly visible from hydrogen-like and helium-like ions for a range of abundant elements including the Ly$\alpha$ lines of {\fexxvi}, {\sixiv}, {\mgxii}, {\nex}, and {\oviii}, and the He-like complexes of {\mgxi}, {\neix}, and {\ovii}.  There is also a bright {\fekalpha} fluorescence line.  A narrow radiative recombination continuum (RRC) from {\neix} is detected indicating that photoionization followed by recombination is the dominant excitation mechanism \citep{LP96}.  We find no obvious absorption features in the spectra.  

\subsection{Phase Dependence}
\label{subsec:phase}

\begin{figure} 
      \centerline{\psfig{figure=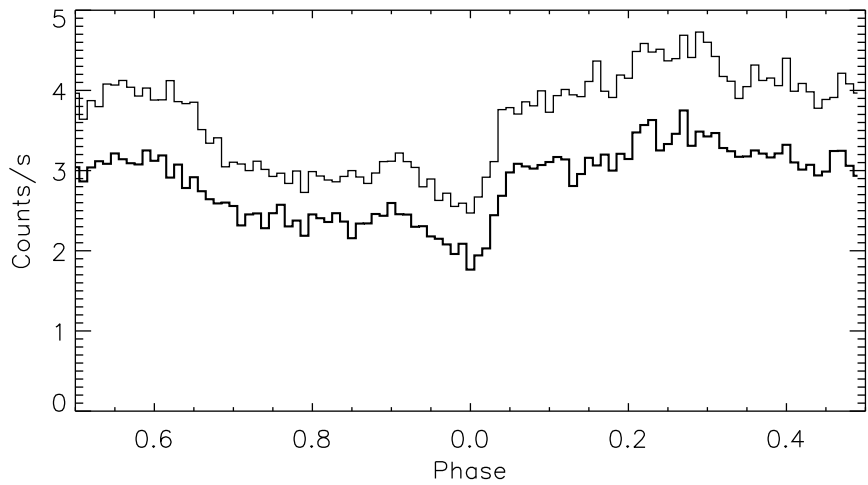,angle=0,height=3.5in}}
      \vspace{-0.0cm}
      \caption{Lightcurve for the HETGS observation in $\Delta \phi=0.01$ phase bins.  The thin and thick histograms show the MEG ($m=\pm 1$) and HEG ($m=\pm 1$) lightcurves, respectively. } 
      \label{FigLightcurve}
   \end{figure}

\begin{figure} 
      \centerline{\psfig{figure=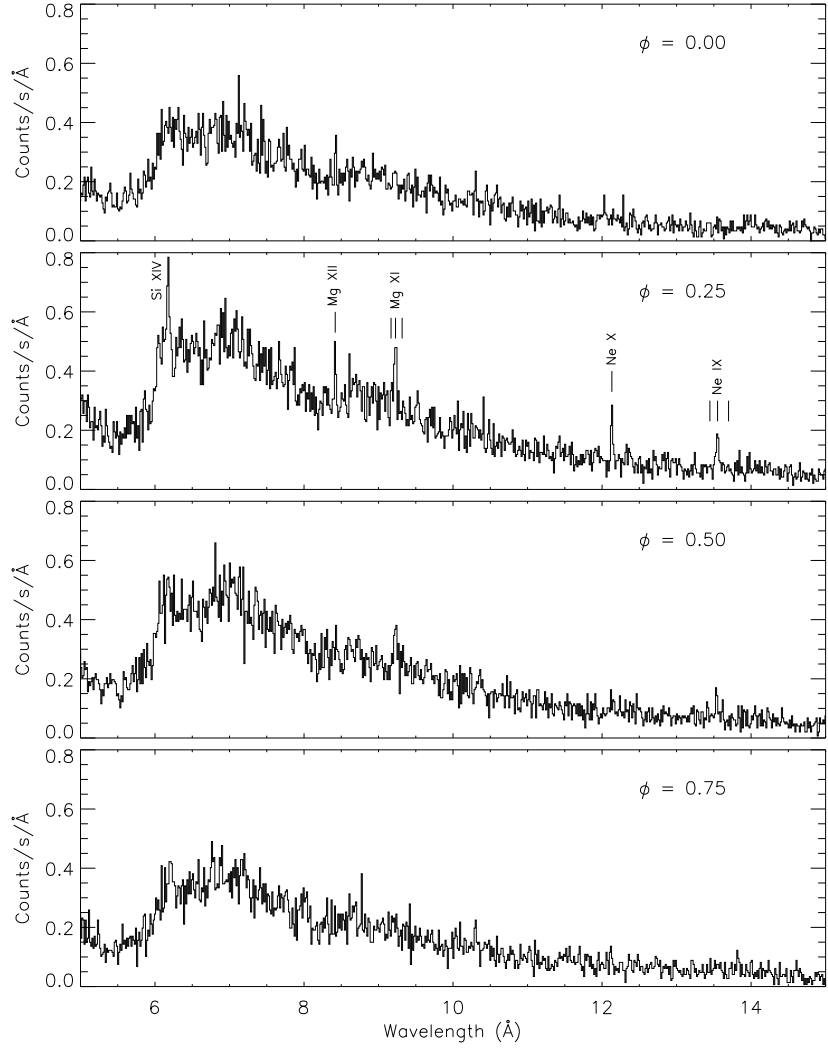,angle=0,height=6.2in}}
      \vspace{-0.5cm}
      \caption{MEG ($m=\pm 1$) spectrum of 4U~1822$-$37 for four phase bins centered on $\phi=0.0, 0.25, 0.50, 0.75$.  } 
      \label{FigMEGVar}
   \end{figure}

\begin{figure} 
      \centerline{\psfig{figure=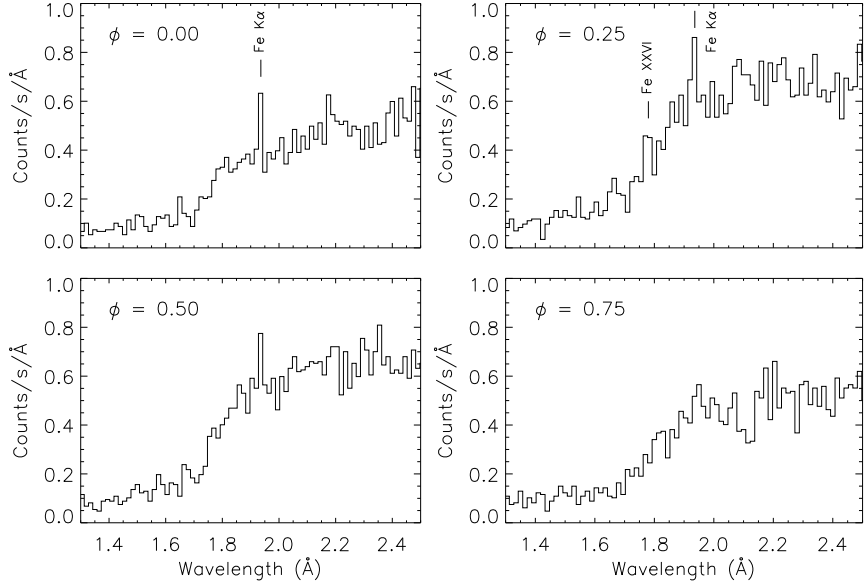,angle=0,height=3.2in}}
      \vspace{-0.0cm}
      \caption{HEG ($m=\pm1$) spectrum of 4U~1822$-$37 for the four phase bins.} 
      \label{FigHEGVar}
   \end{figure}

The observed spectral features are highly phase dependent.  We divided the folded lightcurve (see Figure ~\ref{FigLightcurve}) into four phase bins centered on $\phi=0.00, 0.25, 0.50$, and $0.75$ and collected the spectra for each bin.  In Figure ~\ref{FigMEGVar} we show the MEG spectrum in the wavelength range of $5$ to $15\, {\rm \AA}$ for the four phase bins.  In
Figure ~\ref{FigHEGVar} we show the HEG spectrum in the wavelength range of $1.3$ to $2.5\, {\rm \AA}$.  The {\nex}, {\neix}, {\mgxii}, {\mgxi}, {\sixiv}, and {\fexxvi} emission lines are bright during phase $\phi=0.25$.  During phase $\phi=0.50$ these lines are visible, but considerably weaker.  A weak {\mgxii} emission line is visible during phase $\phi=0.0$, and a weak {\sixiv} line is visible during phase $\phi=0.75$.  The oxygen region in not shown here since the phase-binned count rates are too low for statistically significant analysis.  The phase-dependence of the recombination spectra suggests that the photoionized material comes from a localized region on the disk.  

The recombination lines show a small phase-dependent Doppler shift.  It is difficult to determine precisely the centroid and width of the line emission, given the poor statistics in the phase-binned spectra.  However, during phase $\phi=0.25$ the emission lines are consistent with $v_{\rm Doppler}=(0\pm 150)\, {\rm km/s}$.  During phase $\phi=0.50$ the emission lines show a slight blue shift with $v_{\rm Doppler}=(280 \pm 150)\, {\rm km/s}$.  We see no velocity broadening of the lines with an upper-limit of $\Delta v \leq 300\, {\rm km/s}$. 

The {\fekalpha} line has a very different phase-dependence.  The line is bright during phases $\phi=0.0$, $0.25$, and $0.50$.  During phase $\phi=0.75$ the line flux decreases noticably.  At all phases the line centroid is consistent with $1.935\, {\rm \AA}$, corresponding to transitions in cold, near-neutral material.  We measure an upper limit to the line broadening of $\Delta v \leq 1500\, {\rm km/s}$. 

\subsection{Helium-like Series}
\label{subsec:helike}

\begin{figure} 
      \centerline{\psfig{figure=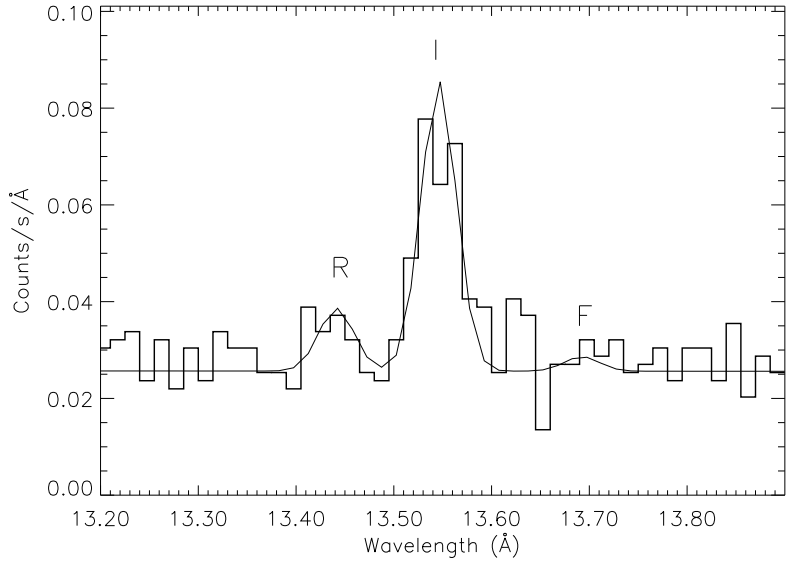,angle=0,height=3.2in}}
      \vspace{-0.0cm}
      \caption[]{Ne~{\small IX} complex in the MEG ($m=\pm1$) spectrum.  The resonance (R), intercombination (I), and forbidden (F) lines are labeled. } 
      \label{FigTriplet}
   \end{figure}

The He-like series of {\ovii}, {\neix}, and {\mgxi} exhibit bright intercombination lines, weak resonance lines and essentially no forbidden lines (see Figure ~\ref{FigTriplet} and the insert on Figure ~\ref{FigSpect}).  The ratio of the intercombination and forbidden emission to the resonance emission, $G=(i+f)/r$, is consistent with a purely recombining plasma;  For {\ovii} and {\neix}, where the lines are cleanly resolved, the measured ratio is $G=4.6 \pm 1.8$ for \ovii and $4.7 \pm 1.5$ for \neix, while the predicted ratios for an optically thin photoionized plasma are $4.0$ and $3.7$ respectively (calculated using HULLAC; \citet{BarShalom}).  The {\mgxi} series lies in the middle of a complex instrumental absorption feature making it difficult to quantify the line ratios.  An enhanced intercombination line is often indicative of a high-density plasma.  However, in 4U~1822$-$37, where there is a significant UV flux \citep[$3\times 10^{-11}\, {\rm ergs/s/cm^{2}}$ from $1200$ to $6400\, {\rm \AA}$:][]{MasonCordovaA}, photoexcitation from the $2^{3}S$ level to the $2^{3}P$ level competes with the spontaneous decay of the $2^{3}S$ line to the ground state.  Using the UV flux reported in \citet{MasonCordovaA} we calculated the photoexcitation rate according to the procedure described in \citet{Kahn2001}.  We calculated a dilution factor of $\sim 0.8$ for the source flux based on the relative geometry of the corona and the accretion disk.  For {\ovii} and {\neix} we find the UV photoexcitation rate to be a factor of $\sim 4 \times 10^{6}$ and $\sim 4 \times 10^{5}$ larger than the rate of radiative decay.  The ratios in the He-like series are therefore insensitive to the plasma density.        

\section{Discussion}
\label{sec:discussion}

\begin{figure} 
      \centerline{\psfig{figure=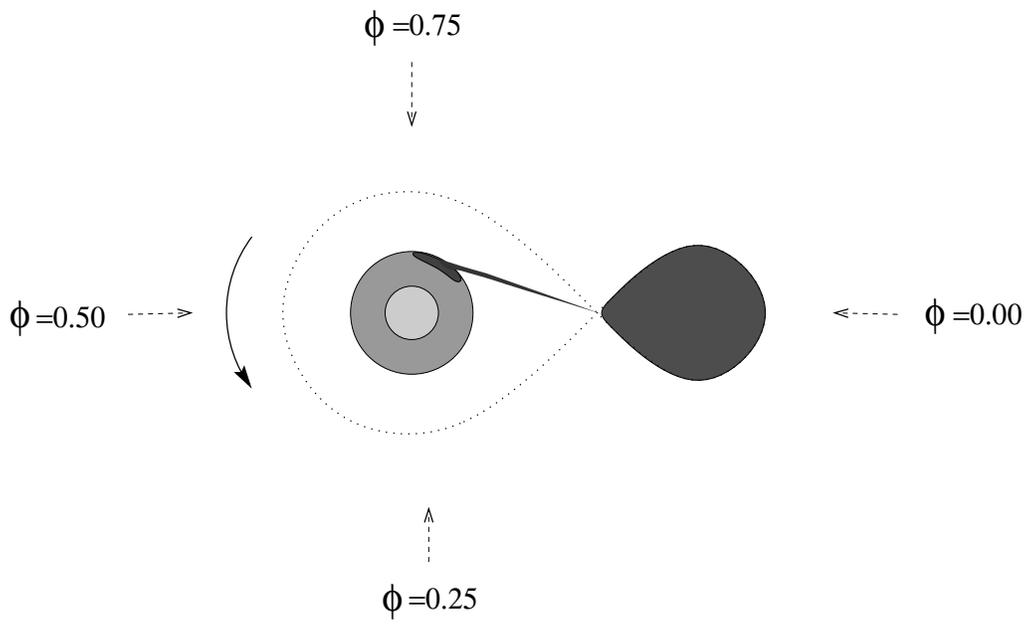,angle=-90,height=3.2in}}
      \vspace{0.5cm}
      \caption{Illustration of the inferred geometry in the 4U~1822$-$37 system.  This illustrates the phase-dependent obscuration of the central source by the bulge, as well as the phase-varying aspect of the irradiated bulge surface. }
      \label{FigCartoon}
   \end{figure}

These spectroscopic features can be used to locate and describe the line emitting regions in 4U~1822$-$37.  The phase-dependence of the recombination line intensity and the lack of absorption features are consistent with emission from an X-ray illuminated, optically thick region on the disk localized between binary phases $\phi=0.75$ and $\phi=1.00$.  During phase $\phi=0.25$ and part of $\phi=0.50$, we see emission lines from the X-ray illuminated inside edge of the material.  During phases $\phi=0.75$ and $\phi=0.00$ this emission region is essentially blocked from view.  This is illustrated in Figure ~\ref{FigCartoon}.  The inferred geometry is insensitive to our choice of phase bins.  The observed velocity structure in the recombination lines is consistent with this scenario;  emission lines from material localized near the edge of the disk at phase $\phi=0.85$ will show very little velocity broadening, but will appear Doppler-shifted by the vector sum of the orbital and Keplerian velocities.  At phase $\phi=0.25$ the lines will appear blue-shifted by $v_{\rm Doppler} \sim 140\, {\rm km/s}$ and at phase $\phi=0.50$ blue-shifted by $v_{\rm Doppler} \sim 360\, {\rm km/s}$.    

The phase location of the emission region is consistent with the point of impact of the accretion stream and suggests that we are seeing line emission from the bulge that is expected to form in the shock-heated, colliding material \citep{Shu,Livio}.  If the material is optically-thick, the bulge is expected to extend above the disk and downstream along the edge of the disk as it adiabatically cools \citep{ArmitageLivio}.  Viewed at high inclination angles, this would produce the wide shallow dip observed in the HETGS lightcurves and in those from previous observations.  

We can estimate the angular size of the material from the flux in the line emission by assuming that the photons above the ionization edge that intercept the material are absorbed and ultimately reemitted as recombination line photons.  Expressing this in terms of the number flux, as $N_{\rm continuum}*\Omega/(4\pi)=N_{\rm absorbed}=N_{\rm recombined}$, and $N_{\rm recombined}*\eta_{\rm line}=N_{\rm line}$, where $\eta$ is the efficiency of cascade through the line we can calculate the solid angle subtended, $\Omega$.  Using the spectrum from the $\phi=0.25$ phase bin we find a solid angle of $\Omega/(4\pi)=(1.8 \pm 0.7) \times 10^{-2}$.  For the 4U~1822$-$37 geometry this corresponds to an illuminated surface area of $(1.0\pm 0.4) \times 10^{21}\, {\rm cm^{2}}$, or a vertical scale height of $\sim (3.3\pm 0.7) \times10^{10}\, {\rm cm}$.

The presence of a bulge at the stream impact point in 4U~1822$-$37 was first suggested by \citet{Mason} to explain the sinusoidal modulation in the optical lightcurve.  Using detailed modeling of the X-ray, UV, optical, and IR lightcurves \citet{WhiteHolt}, \citet{MasonCordovaB}, and \citet{HellierMason} all inferred a phase-dependent vertical structure along the edge of the disk that is dominated by a bulge around phase $\phi=0.85$, which extends from $\phi=0.70$ to $\phi=0.0$ with a maximum vertical height between $0.6$ and $1.6 \times 10^{10}\, {\rm cm}$ depending on the model.  This is consistent with the phase location and the vertical height that we measure here to within a factor of two.  

We can use the details of the HETGS spectra to further constrain the physical parameters of the emitting material.  The width of the RRC feature is a direct measure of the plasma temperature.  For the cleanly resolved {\neix} RRC we find a temperature of $13\pm 7\, {\rm eV} = (1.5\pm 0.8) \times 10^{5}\, {\rm K}$.  We can estimate the plasma density of the emitting material using the emission measure, ${\rm EM}=\int{dV n_{\rm e}^2}$.  The emission measure is determined by dividing the line luminosity by the line powers, which are calculated using HULLAC \citep{BarShalom}.  For all ions except oxygen, we measure ${\rm EM} \sim 5 \times 10^{56}\, (D/2.5\, {\rm kpc})^{2}\, {\rm cm^{-3}}$.  For {\oviii} and {\ovii} we measure ${\rm EM} \sim 2.5 \times 10^{55}\, (D/2.5\, {\rm kpc})^{2}\, {\rm cm^{-3}}$, which could imply a relative underabundance of oxygen by a factor of $\sim 20$.  Estimating the volume by taking the vertical height that we infer here, the azimuthal extent from previous modeling of the lightcurves, and assuming that the emission measure is dominated by material within a Thompson depth of $\tau=1$, we find an average electron density of $\sim 1 \times 10^{11}\, {\rm cm^{-3}}$.  Such a high density would have been discernable in the He-like line ratios if the UV photoexcitation had not saturated the transitions.   

The persistence of the {\fekalpha} fluorescence emission throughout the binary phases with a decrease in the intensity only during phase $\phi=0.75$ suggests that the fluorescing material comes from an extended inner region of the system, which is visible at all times, but partially blocked from view at that point by the bulge.  The width of the line constrains the radial location of the emitting material.  From the measured upper-limit to the velocity broadening we find a lower limit for the inner radius of $R \geq 1\times 10^{10}\, {\rm cm}$.  According to models for the formation of the corona, it should be highly ionized and optically thick \citep{WhiteHolt}.  The observed {\fekalpha} line cannot originate in such a corona or else it would have a higher energy centroid and would appear broadened by Compton scattering in the hot ($\sim 1.8\, {\rm keV}$) coronal material.  

We can estimate the relative geometry of the fluorescing material and the illuminating central source using the flux in the fluorescence emission line.  Calculating the illuminated solid angle using the same assumptions as before, but now with the fluorescence yield of neutral iron, we find $\Omega/(2\pi)=0.11\pm 0.3$.  This is much larger than expected for an embedded point source illuminating a disk, unless the edge of the disk flares to scale heights comparable to the bulge height.  In order to illuminate such a large solid angle, the central source must be extended.  However, the fluorescing material must be even more extended; an extended source laying over an illuminated region of comparable size produces a solid angle of $\Omega\sim 2\pi$, which is much larger than we observe here.  The cold clumps formed by instabilities in an accretion stream that skims over the disk surface \citep{FKL} might provide sufficient solid angle.  Alternatively, the disk around an embedded spherical central source subtends a solid angle of $\Omega/(2\pi)\leq 0.25$ \citep{Halpern}.  A covering fraction of $0.11$ occurs when the inner radius of the illuminated material is about twice as large as the radius of the spherical source.  In 4U~1822$-$37 this would give a lower-limit to the radius of the corona of $R_{\rm corona} \geq 5 \times 10^{9}\, {\rm cm}$, which is only a factor of $\sim 5$ smaller than the radius determined by modeling the lightcurve.


\acknowledgments
Support for the work at Columbia University was provided by the NASA through Chandra Award number SAO-G00-1099X issued by the Chandra X-ray Observatory Center under contract NAS8-39073.  Work at LLNL was performed under the auspices of the U.S. Department of Energy by the University of California Lawrence Livermore National Laboratory under contract No. W-7405-Eng-48.

\clearpage


\end{document}